# Sputter-Induced Cross-Contaminations in Analytical AES and XPS Instrumentation: Utilization of the effect for the In-situ Deposition of Ultrathin Functional Layers

Uwe Scheithauer

*Unterhaching*

Phone: +49-(0)89-63644143

email: scht.uhg@googlemail.com

## Keywords:

## Abstract

Cross contaminations are observed on sample surfaces by AES and XPS, if multiple samples are mounted on one sample holder and a neighbouring sample was sputter depth profiled. During sputter depth profiling sputtered material is deposited on inner surfaces of the instrument. In a secondary sputter process, which is due to species leaving the primary sputter target with higher kinetic energy, the previously deposited material is transported from the inner surfaces to the other samples mounted on the sample holder.

This Reflective Sputtering is utilized to deposit ultra-thin layers on sample surfaces for XPS binding energy referencing purpose and to build up ultra-thin conductive layers to enable AES measurements on insulating samples.

# Introduction

Sputter induced cross-contaminations, also denoted as memory effect, have been discussed in context with *secondary ion mass spectrometry* (SIMS) depth profile measurements *[1, 2]*. This paper reports about sputter-induced cross-contaminations observed by *Auger Electron Spectroscopy* (AES) and *X-ray Photoelectron Spectroscopy* (XPS). Since the elemental detections limits of AES and XPS are ~ 0.1 at% in favourable cases and the SIMS detection limits are below the 1 ppm level, it is not expected to see a memory effect in AES or XPS depth profile measurements. But if in modern equipment multiple samples are mounted at the same time on a sample holder, cross-contaminations are observed on sample surfaces if other, neighbouring samples have been sputter depth profiled before. This cross-contamination effect is investigated and simulated by an experiment. The experiment explains that the effect is a result of material deposited on inner instrument surfaces, which is sputtered by high-energy particles emitted by the primary sputter depth profiling. But this effect of Reflective Sputtering is not only troublesome in producing misleading analysis results. It is shown how it is converted into a powerful tool to deposit ultra-thin conductive layers which enable AES measurements on insulating samples or to deposit ultra-thin metal layers on samples for XPS binding energy referencing purpose.

# Experimental

For the measurements presented here an AES microprobe PHI 680 and an XPS microprobe Quantum 2000 were used, respectively. Both instruments are manufactured by Physical Electronics.

The PHI 680 Auger microprobe, an instrument with a Schottky thermal field emitter, has a lateral resolution of ~ 15…30 nm at optimum *[3]*. The Auger microprobe is equipped with differentially pumped $Ar^+$ ion sputter gun. If a 30° sample holder tilt is used as here, the primary electron beam hits the surface of a flat sample under an angle of 30° relative to the surface normal. The $Ar^+$ ion impact angle is ~ 45° relative to the surface normal in that case.

The Quantum 2000 is an XPS instrument with a focused primary X-ray beam. The spatial resolution of a Quantum 2000 XPS microprobe is achieved by the combination of a fine-focused electron beam generating the X-rays on a water cooled Al anode and an elliptical mirror quartz monochromator, which monochromatizes and refocuses the X-rays to the sample surface *[1-6]*. By scanning the electron beam electrostatically across the Al anode the X-ray beam scans across the sample. Using a rastered X-ray beam, sample features are localized by X-ray beam induced *secondary electron* (SE) images. The best focused X-ray beam has a diameter of ~ 10 µm. Beam diameters up to 200 µm are selectable. Details of the quantitative lateral resolution are discussed elsewhere *[7]*. For flat mounted samples as used here in a Quantum 2000, the incoming X-rays are parallel to the surface normal. In this geometrical situation, the mean geometrical energy analyser take off axis and the differentially pumped $Ar^+$ ion gun, which is used for charge neutralization, sputter cleaning and depth profiling of the samples, are oriented ~45° relative to the sample surface normal.

The differentially pumped Ar ion guns of both systems have an identical design. To avoid neutral Ar atoms hitting on the sample surface it has an electrostatic beam deflection of ~ 5°. The ion gun has a floating column so that it can supply $Ar^+$ ions with a kinetic energy of a few *electron volts* (eV) and a current high enough to provide effective compensation of sample charging in XPS and to enable AES measurements on electrically conductive features in insulating surroundings, respectively *[8]*.

The PHI software Multipak 6.1 was used for the AES and XPS data evaluation. In case of quantification of measured peak intensities it uses the simplified model, that all detected elements are distributed homogeneously within the analysed volume. This volume is defined by the analysis area and the information depth of electron spectroscopies, which is derived from the mean free path of electrons *[9]*. Using this quantification approach one monolayer on top of a sample quantifies to ~ 10 … 30 at% depended on the samples details.

## Sputter Induced Cross Contamination

Sputter induced cross contamination became evident by unexpected measurement results. One of these unexpected results was from the analysis of a Si wafer surface. It is discussed and the effect is investigated systematically.

### Unexpected XPS Analysis Results of a Si Wafer Surface

The aim of the analysis was to inspect a Si wafer surface after a cleaning process at the beginning of the front end wafer processing. Pieces of a Si wafer were analysed by XPS. A small Au signal was detected at the Si wafer surface 'as received' (fig. 1). Using tabulated sensitivity factors, one can estimate the amount of Au to be a few percents of an atomic monolayer. Under usual circumstances metal contamination in this early stage of front end wafer processing is not present or expected. Therefore this result has to be checked very carefully regarding any measurement artefact.

In fig. 1 the insert shows the Quantum 2000 sample holder with all samples mounted. Besides two pieces of the cleaned Si wafer (light grey) two pieces of fully processed wafers (dark grey) with an Au (350 nm) / Ti (700 nm) / Al (~ 400 …500 nm) back-side metallization are mounted. For different reasons depth profiles of the two back-side metallizations were measured first before analysing the cleaned Si wafer surfaces. When the analysis of the cleaned Si wafer surface was repeated on new pieces of the same cleaned Si wafers, Au was no longer detected. So it was proven that the Au signal measured on the surface of the cleaned Si wafer before was a cross contamination due to the sputter depth profiling of the Au / Ti / Al metallizations.

Even though the sputter depth profiles of the backside metallization was done through the whole Au / Ti / Al metallization layer stack, only Au was detected on the surface of the cleaned Si wafer. This is explained by the relative ionisation cross sections, which are 0.537, 7.91 and 17.12 for Al2p, Ti3p and Au4f, respectively *[10]*. These values correspond to the over-all sensitivity factors of an X-ray microprobe Quantum 2000 used by the data evaluation software Multipak 6.1, which are approximately 0.26, 2.1 and 6.8 for Al2p, Ti3p and Au4f. Hence Au is detected more effectively and the other elements are not detected in the measured spectrum (-> fig. 1).

## Systematic investigation of the sputter induced cross contamination in an XPS microprobe Quantum 2000

Fig. 2 summarizes the result of a systematic investigation of the sputter-induced cross-contamination. As sputter target again the Au / Ti / Al backside metallization was used. On both sides, pieces of a clean Si wafer were mounted. The metallization stack was sputtered by 2 kV $Ar^+$ ions with an impact angle of ~ 45° relative to the surface normal. The sputtering was interrupted after 5, 10 and 30 minutes and ended after 130 minutes. After 130 minutes the whole metallization system was sputtered away. At each interruption of sputtering and at the end, the Au cross-contamination on the Si pieces was measured as function of the relative position with regard to the sputter crater. After a sputter time of 130 minutes, Ti was detected too. Fig. 2 shows that the Au coverage on the Si increases with increasing sputter time. Furthermore, it shows that relative to the incoming $Ar^+$ ions, the coverage in forward direction is higher than in backward direction.

# Analysis of inner surfaces of the instrument

During sputter depth profiling the sputter material is emitted into the half-sphere above the sample and is deposited everywhere on the inner surfaces of the instrument. The drawing in the upper part of fig. 3 shows the geometrical situation in an XPS microprobe Quantum 2000. In the half sphere above the sample a lot of mechanical instrument components are present. These devices are the Al foil window of the X-ray source and its mechanical mounting, parts of the $Ar^+$ ion gun and the electron energy analyser entrance. Together with the electron neutraliser, these devices are fixed to a solid metal block. The Al foil window has to be replaced from time to time due to an X-ray intensity decrease. This way one has easy access to the inner surface of the instrument, which is located very close to the sample. The lower part of fig. 3 shows a used Al foil. In the exposed area the colour has changed. Fig. 4 shows the results of an XPS depth profile measurement of the material deposited on the foil. As depicted, a lot of elements are detected in the contamination layer. These elements and their depth distribution record the history of past sputter depth profile measurements. With the ion dose used to remove the contamination layer, ~ 35 nm of the reference material $SiO_2$ could be removed. Since the sputter rates of metals are higher than the $SiO_2$ sputter rate *[11, 12]*, presumably the geometrical thickness of the layer is greater. And of course, this deposited material explains the primary X-ray intensity decrease.

# Principle of Reflective Sputtering

A Reflective Sputtering experiment was designed on the bases of the previous results. Fig. 5 shows the principle of the experiment. On the sample holder we have the primary sputter target and a piece of clean Si wafer. Above both a Si reflector, again a piece of clean Si wafer, is mounted in a distance of a few millimetres. The primary target is sputtered by $Ar^+$ ions. I.e. sputtered material is deposited on the Si reflector. This material, which was deposited previously on the reflector, is sputtered by primary $Ar^+$ ions, which are reflected, and by sputtered species of higher kinetic energy, which are emitted from the primary target *[13, 14]*. Some of this secondarily sputtered material from the reflector is deposited on the piece of Si wafer. In summary, this way material is transported from the primary target to the piece of Si wafer by Reflective Sputtering.

The Reflective Sputtering experiment was done using an Auger microprobe PHI 680. In a first sputter step, a primary Ag target was sputtered with 2 kV $Ar^+$ ions for sputter time $t_0$. Then the primary target was replaced by Au and only now a clean Si wafer piece was mounted below the Si reflector. In a second sputter step, the Au target was sputtered too, for the same time $t_0$.

The surface coverage with Ag and Au deposited on the Si reflector and the Si wafer was measured in an XPS microprobe Quantum 2000, because the detection sensitivity for both elements is higher using XPS instead of AES. The results are presented in fig. 6. The measured Ag and Au XPS signal intensities are plotted as function of the distance from the forward edge of the Si reflector and Si wafer, respectively. On both samples the Ag and Au intensity decreases with the distance from the edge. On the Si reflector, signal intensities higher by a factor of ~5 were detected. The metal layer thickness is estimated to be ~2…3 monolayers on the reflector and ~ 0.5 monolayer on the Si wafer at maximum.

These results show that Ag from a pre-coated Si reflector and Au are deposited on the Si wafer during the Reflective Sputtering of Au. In summary, this experiment is a perfect simulation of the observed sputter-induced cross-contamination process.

# Application of Reflective Sputtering

In the daily analytical work misleading results, which are due to sputter induced cross contaminations, are avoidable. Therefore, on a sample holder with multiple samples, all surfaces 'as received' must be measured before any sputter depth profiling will be done. But Reflective Sputtering is not only a cumbersome concomitant phenomenon of sputter depth profiling. It rather can be converted into a powerful tool which introduces new approaches to electron spectroscopy measurements.

### Use of an ultrathin conductive layer for AES measurements of an insulating sample

A recent article summarizes the approaches how to analyze insulators with AES *[15]*. Using reflective sputtering adds a new method to solve this problem. The basic concept is the same one as is commonly used in electron microscopy. On top of a non-conductive sample a thin conductive layer is deposited. The thickness of the ultrathin conductive layer on top is restricted to a few monolayers only, since roughly only the topmost 10 monolayers contribute to the signal of an Auger measurement and the contribution of each monolayer decreases exponential with the depth *[9, 16]*.

To demonstrate this approach, on the Auger instruments sample holder a Cu foil was mounted beside an $Al_2O_3$ ceramic sample. Then the Cu foil was sputtered by $Ar^+$ ions. Using the inner surfaces of the Auger instrument as reflector, in-situ an ultra-thin Cu layer was deposited on the surface of the ceramic sample. As defined by the incoming $Ar^+$ ion beam, the $Al_2O_3$ ceramic sample is mounted in forward direction relative to the Cu foil. As seen earlier, this gives higher deposition rates.

Fig. 7 shows a comparison of the uncoated $Al_2O_3$ ceramic sample in the upper part with a coated sample in the lower part. The SE image of the uncoated sample depicts the typical behaviour of sample charging. Drastic changes of SE intensity are observed from the left to the right and from line to line according to the raster scan directions of the SE image. The Auger spectrum is completely distorted and can not be interpreted. The SE image of the coated sample is fine and has a sub-μm resolution. The Auger spectrum shows no serious distortions. Only between 145 and 195 eV does the spectrum show some distortions. Besides Cu, the elements C, O, Al and Si are detected. This result is comparable to an XPS measurement of the same sample. Using quantification as described earlier, one can estimate the Cu layer thickness to be ~1…2 monolayers.

First, from the measurement we see that ultra-thin conductive metal layers make Auger measurements of insulators possible. But the measurement on insulating samples is something which modern XPS equipment does without any effort. So, second, it has to be pointed out that the metallization technique using Reflective Sputtering enables measurements on insulating samples with the much higher lateral resolution of an Auger instrument.

This approach of in-situ metal deposition may fail depending on the interaction of the insulators surface with the deposited metal. If, for instance, the deposited metal atoms are mobile on the insulators surface, they may form three-dimensional clusters rather than a conductive film. In detail, such processes are dependent on the insulators bulk material, the insulators surface contamination, the kind of metal used and the operating temperature. In experiments using a thick insulating $SiO_2$ layer, the deposition of several monolayers of Cu at room temperature did not form a conductive layer, for instance. In any case, besides some unsuccessful attempts, the reflective sputter deposition is a useful approach to measure insulators using the lateral resolution of Auger instrumentation.

**Ultra-thin Metal Layers for Energy-Referencing Purpose in XPS Measurements**

For the energy scale calibration of XPS instruments with a monochromatic $Al_{k\alpha}$ X-ray source the Au4f7/2, Ag3d5/2 and Cu2p3/2 peaks are used as binding energy references *[17]*. Reflective Sputtering was utilized for this experiment with an XPS microprobe Quantum 2000 to deposit Cu and Au on the surface of an $Al_2O_3$ ceramic sample for energy referencing purpose. The insert and the optical image in fig. 8 illustrates the experimental set-up. Pure Cu and Au foils are mounted close together on the sample holder by a mechanical clamp. Both foils are sputtered simultaneously with a rastered $Ar^+$ ion beam. This way Cu and Au are deposited on the surface of the $Al_2O_3$ ceramic sample via Reflective Sputtering. By the quantification routines of the Multipak software program, the amount of material deposited on the $Al_2O_3$ ceramic sample surface was estimated to be 1.2 at% Au and 0.5 at% Cu. So, on top of the sample we have metal layers within a sub monolayer thickness range.

Fig. 8 shows some peaks of a high energy resolution measurement. Tab. 1 summarizes the results. The measured binding energy difference between Cu3p3/2 and Au4f7/2 is by 0.118 eV smaller than the-value, which is given by the ISO 15472 *[17]*. The results shown in the left binding energy column are the measured values shifted by 3.199 eV. For the results shown in the right column a linear function was used, setting the Cu2p3/2 and Au4f7/2 binding energies to the reference values of 932.62 and 83.96, respectively *[17]*. The XPS microprobe used here is operated with a calibrated energy scale *[18]*. Over long periods a precision of ±0.3 eV is achieved on conductive specimens. From the elaborated effort done within the framework of energy scale calibration, it is known that the precision of one single binding energy measurement is estimated to be approximately ±0.12 eV. Taking this precision into account, we find both energy scale corrections are valid.

In the past, many attempts have been made to deposit materials on XPS sample for reference purpose. For instance, Au was evaporated in vacuum onto different materials and the Au 4f7/2 was used for energy scale adjustment *[19]*. With increasing amount of Au slight binding energy changes are observed. For the system Au on Ni a lowering of the Au4f7/2 binging energy of ~ 0.3 eV for a coverage below 1 monolayer is reported, for instance *[20]*. To avoid these energy shifts, which may be explained by a surface having only single Au atoms or small Au nanoparticles build from a few atoms on top, colloidal noble metal particles dispersed in a high purity liquid matrix were used in an ex-situ deposition process *[21]*.

In summary and with regard to the achievable precision of a binding energy measurement, every deposition of the reference materials Cu, Ag and Au represents an improvement of peak binding energy estimation. Reflective Sputtering has the main advantage that for this approach the hardware needed, a sputter ion gun, is already available in a typical analytical XPS instrumentation. Since it is an in-situ technique, binding energies can be recalibrated after sputter erosion of a sample simply by Reflective Sputter depositioning of reference materials on this surface.

# Summary

Sputter-induced cross-contamination on sample surfaces are observed with the electron spectroscopy methods AES and XPS. This contaminations are due to prior sputter depth profiling of other samples mounted at the same time on the same sample holder. In detail this cross contamination is described by two processes. Sputtered material from the primary sputter depth profiling is deposited on the surface of mechanical devices which are positioned near the sample. As sputtering continues, in a secondary sputter process previously deposited material is removed from these inner surfaces. The secondary sputter process is attributed to reflected primary sputter ions and sputtered material, provided that both species have a high enough kinetic energy. In summary, the Reflective Sputtering deposits material on surfaces of other samples mounted on the sample holder during sputter depth profiling of one sample. To avoid misleading results in analytical AES and XPS work, all surfaces 'as received' must be measured first before any depth profiling is done.

But Reflective Sputtering is not only a cumbersome process. It can be converted into a powerful tool. Within Auger instrumentation ultra-thin conductive layers can be deposited on insulating samples in order to make possible measurements of insulator surfaces with the lateral resolution of an electron beam. With XPS instrumentation Reflective Sputtering may be utilized to deposit ultrathin metal layers on sample surfaces for binding energy referencing purpose. Both applications are easily realizable by using the depth profile sputter gun and metal foils mounted near to the sample.

## Acknowledgement

All data were measured utilizing an XPS microprobe Quantum 2000 and an AES microprobe PHI 680 installed at Siemens AG, Munich, Germany. I acknowledge the permission of the Siemens AG to use the measurement results here. For many fruitful discussions I would like to express my thanks to my colleagues.

# References

[1] Deline VR (1983) Instrumental cross-contamination in the Cameca IMS-3F secondary ion microscope. Nucl. Instrum. Methods Phys. Res. 218:316-318

[2] Wittmaack K (1985) Experimental and theoretical investigations into the origin of cross-contamination effects observed in a quadrupole-based SIMS instrument. Appl. Phys. A 38:235-252

[3] Physical Electronics Inc. (2001) System Specifications PHI Model 680. Eden Prairie, MN 55344 USA

[4] Iwai H, Oiwa R, Larson PE, Kudo M (1997) Simulation of Energy Distribution for Scanning X-ray Probe. Surf. Interface Anal. 25:202-208

[5] Physical Electronics Inc. (1997) The PHI Quantum 2000: A Scanning ESCA Microprobe. Eden Prairie, MN 55344 USA

[6] Physical Electronics Inc., System Specifications for the PHI Quantum 2000, Eden Prairie, MN 55344 USA (1999)

[7] Scheithauer U (2008) Quantitative Lateral Resolution of a Quantum 2000 X-ray Microprobe. Surf. Interface Anal. 40:706-709

[8] Iwai H, Namba H, Morohashi T, Negri RE, Ogata A, Hoshi T, Owia R (1999) A Study of Charge Compensation for Insulator Samples in AES by Low Energy Ion Beam Irradiation. J. Surf. Anal. 5:161-164

[9] Seah MP, Dench WA (1979) Quantitative electron spectroscopy of surfaces: A standard data base for electron inelastic mean free paths in solids. Surf. Interface Anal. 1:2-11

[10] Scofield JH (1976) Hartree-Slater subshell photoionization cross-sections at 1254 and 1487 eV. J. Electron Spectrosc. Relat. Phenom. 8:129-137

[11] Andresen HH, Bay HL (1981) Sputtering Yield Measurements. In: Behrisch R (Ed.) Sputtering by Particle Bombardment I. Springer, Berlin

[12] Betz G, Wehner GK (1983) Sputtering of Multicomponent Materials. In: Behrisch R (Ed.) Sputtering by Particle Bombardment II. Springer, Berlin

[13] Biersack JP, Eckstein W (1984) Sputtering studies with the Monte Carlo Program TRIM.SP. Appl. Phys. A 34:73-94

[14] Gnaser H (2007) Energy and Angular Distribution of Sputtered Species. In: Behrisch R, Eckstein W (Eds.) Sputtering by Particle Bombardment. Springer, Berlin

[15] Baer DR, Lea AS, Geller JD, Hammond JS, Kover L, Powell CJ, Seah MP, Suzuki M, Watts JF, Wolstenholme J (2010) Approaches to analyzing insulators with Auger electron spectroscopy: Update and overview. J. Electron Spectrosc. Relat. Phenom. 176:80-94

[16] Seah MP (1983) Quantification of AES and XPS. In: Briggs D, Seah MP (Eds.) Practical Surface Analysis by Auger and X-ray Photoelectron Spectroscopy. Wiley, Chichester

[17]  ISO 15472 (2001) Surface chemical analysis - X-ray photoelectron spectrometers - Calibration of energy scales, ISO, Geneva

[18]  Scheithauer U (2012) Long time stability of the energy scale calibration of a Quantum 2000. J. Electron Spectrosc. Relat. Phenom. 184:542-546

[19]  Kohiki S, Oki K (1985) An appraisal of evaporated gold as an energy reference in X-ray photoelectron spectroscopy. J. Electron Spectrosc. Relat. Phenom. 36:105-110

[20]  Steiner P , Hüfner S (1981) Core level binding energy shifts in Ni on Au and Au on Ni overlayers. Solid State Commun. 37:279-283

[21]  Gross T, Richter K, Sonntag H, Unger W (1989) A method for depositing well defined metal particles onto a solid sample suitable for static charge referencing in X-ray photoelectron spectroscopy. J. Electron Spectrosc. Relat. Phenom. 48:7-12

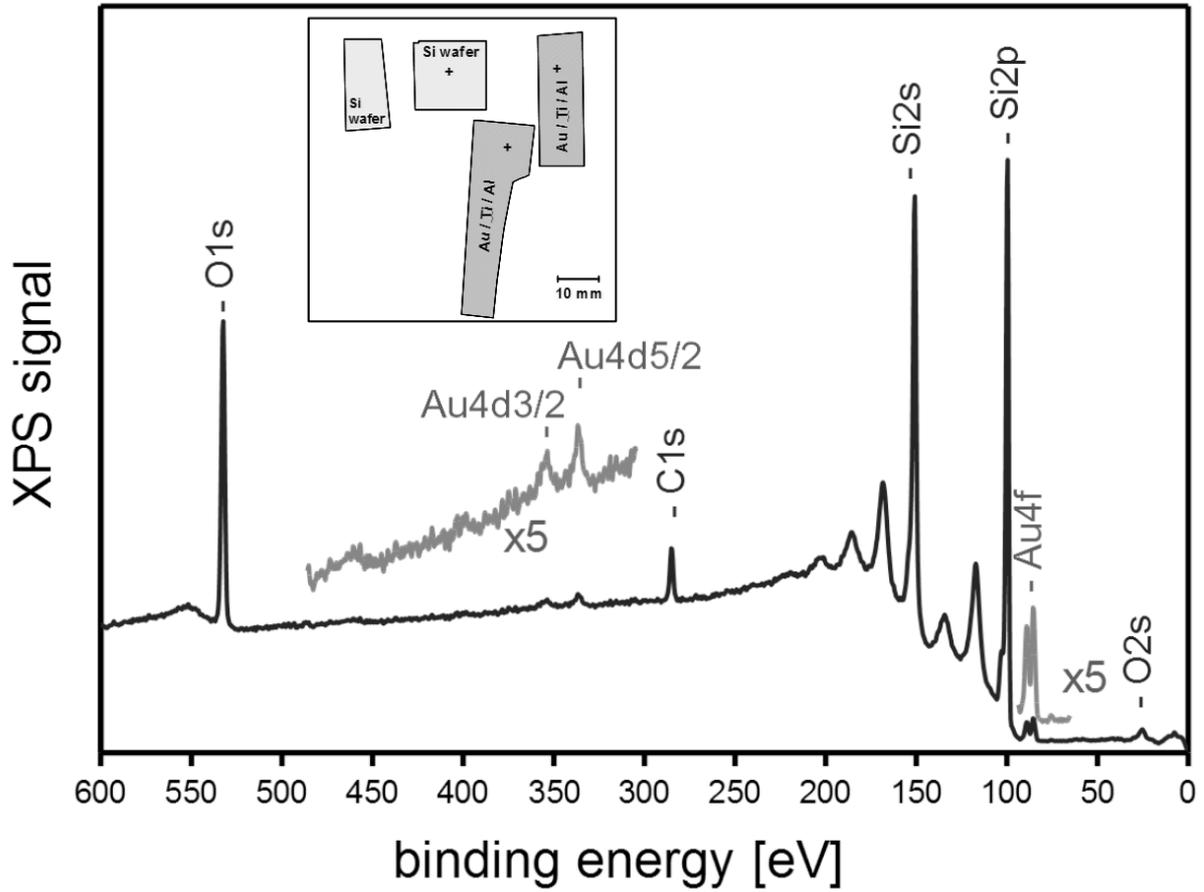

**Fig. 1**: XPS survey scan of a Si wafer surface after a cleaning process
Unexpectedly Au was detected. It is due to redeposition during sputter depth
profiling of an other sample mounted on the sample holder (see the insert).

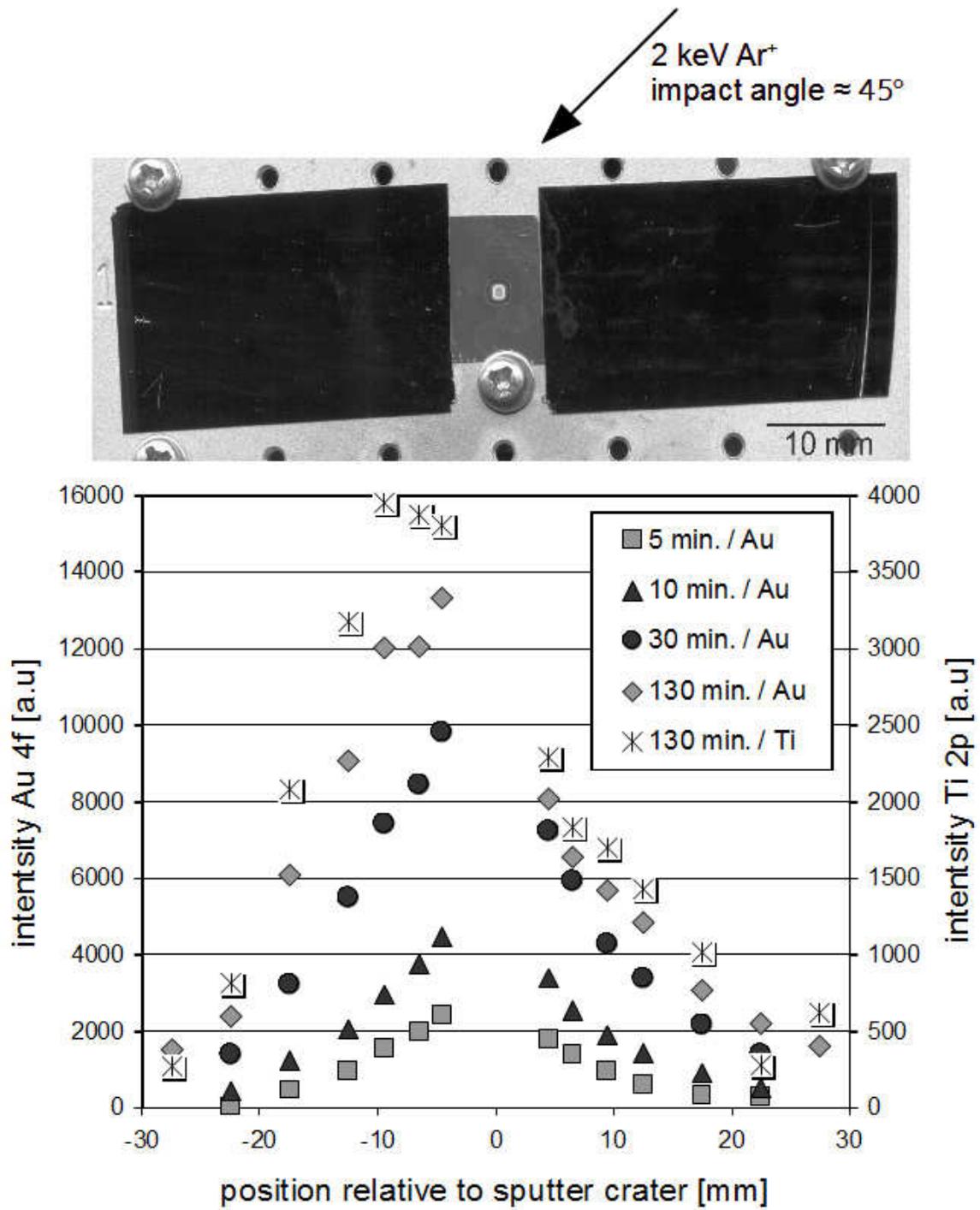

**Fig. 2**: sputter induced material redeposition in an Quantum 2000 XPS microprobe
The graph shows the amount of redeposited material as function of the distance from
the sputter carter and sputter time.

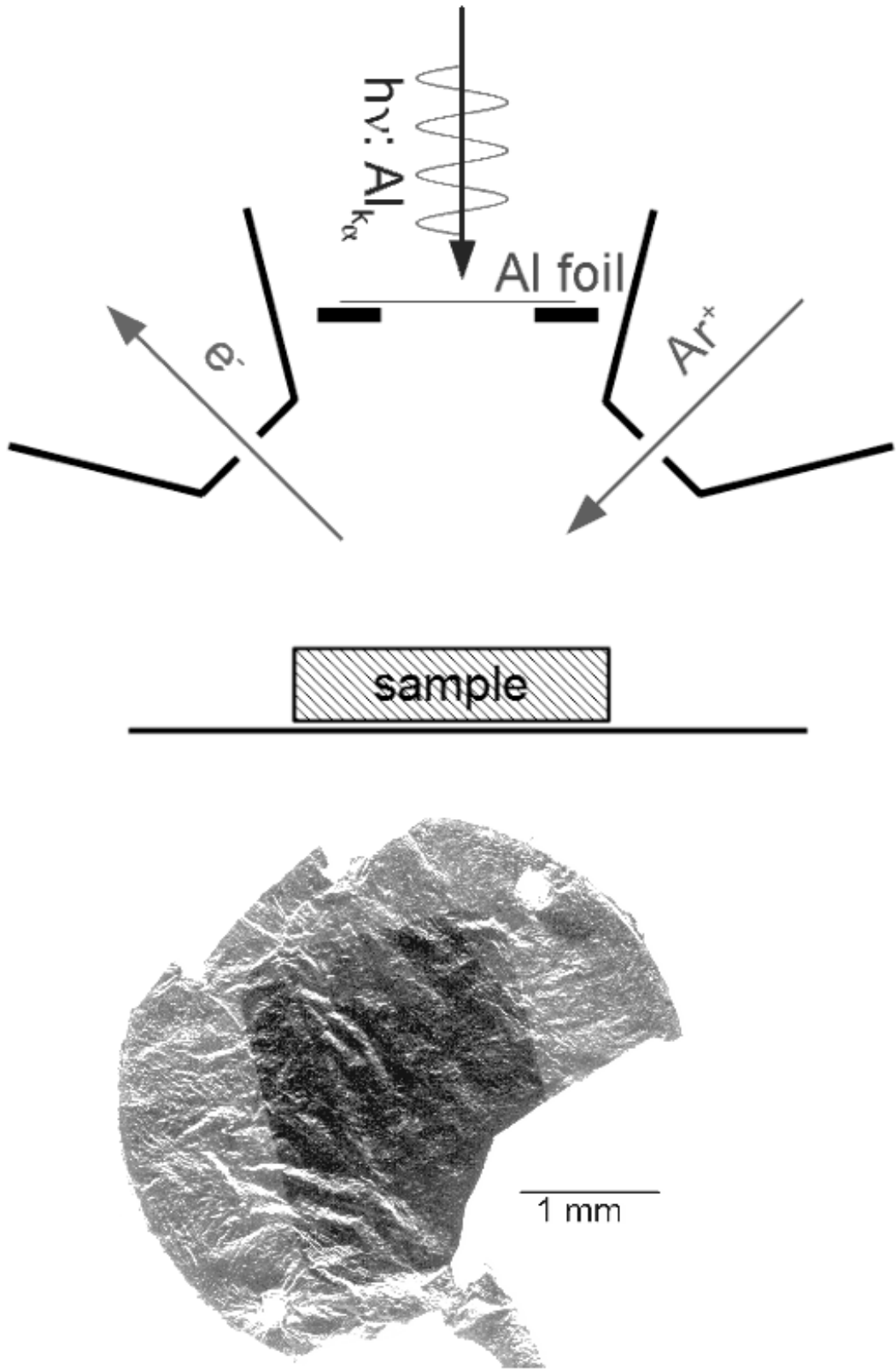

**Fig. 3**: upper part: drawing of the Quantum 2000 sample region with instrument devices mounted in the half sphere above the sample

lower part: Al foil used as separation between X-ray source and sample/rest of the XPS instrument. The exposed area of the foil shows a change in colour.

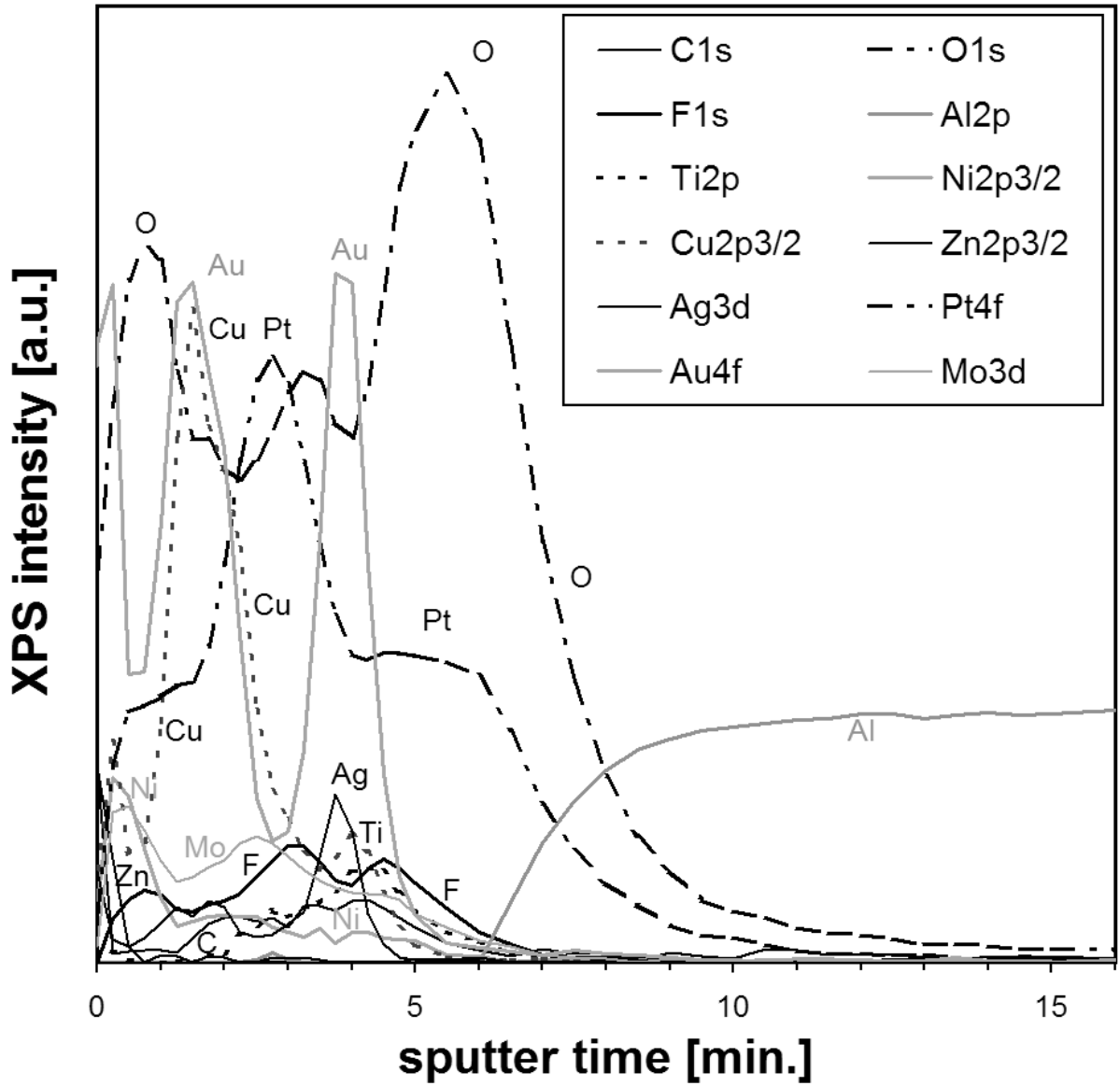

**Fig. 4**:  sputter depth profile of the exposed area of the Al foil X-ray window shows a
contamination layer at the surface

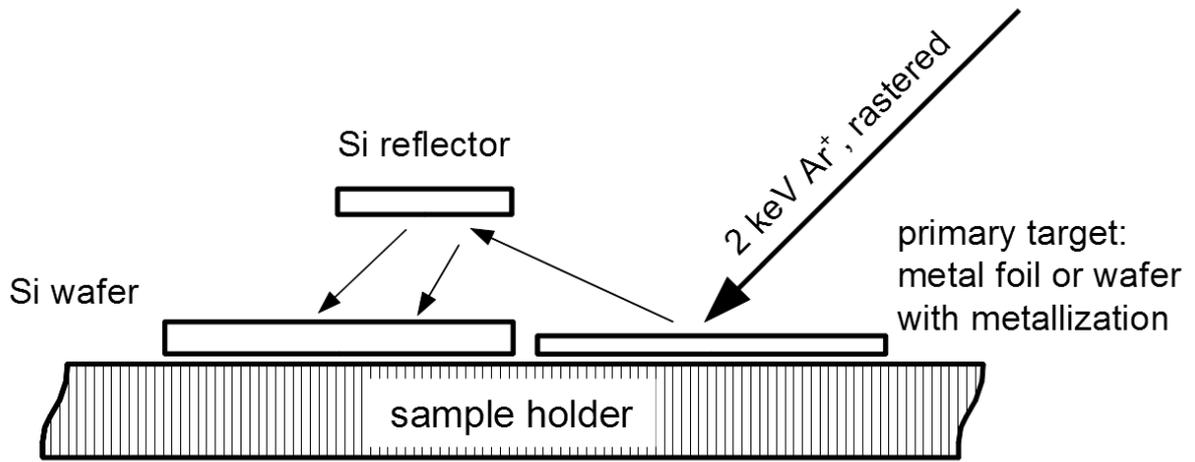

**Fig. 5**:   Reflective Sputtering

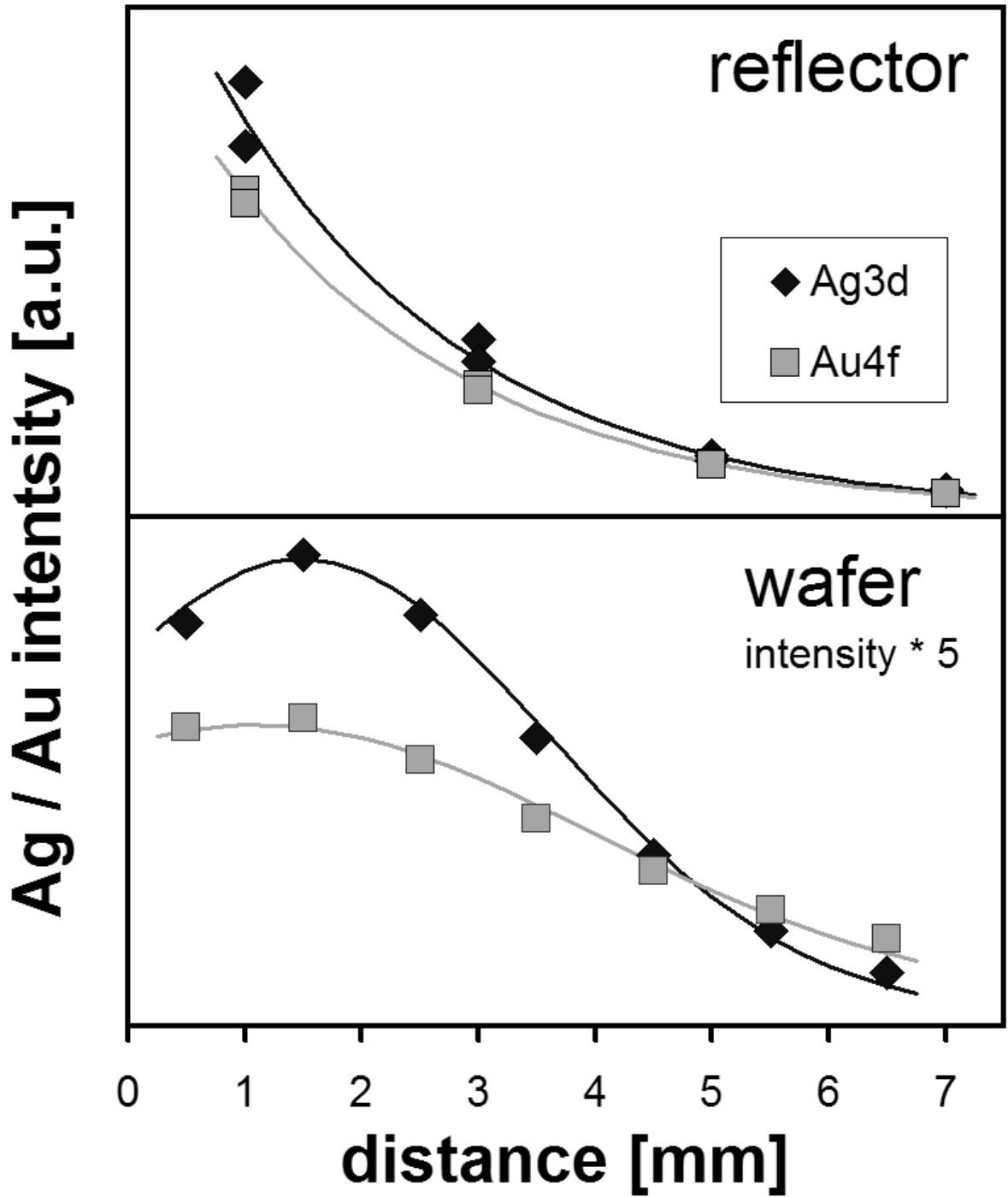

**Fig. 6**:   Au and Ag XPS intensity on a Si reflector and on a Si wafer as function of the distance from the forward edge

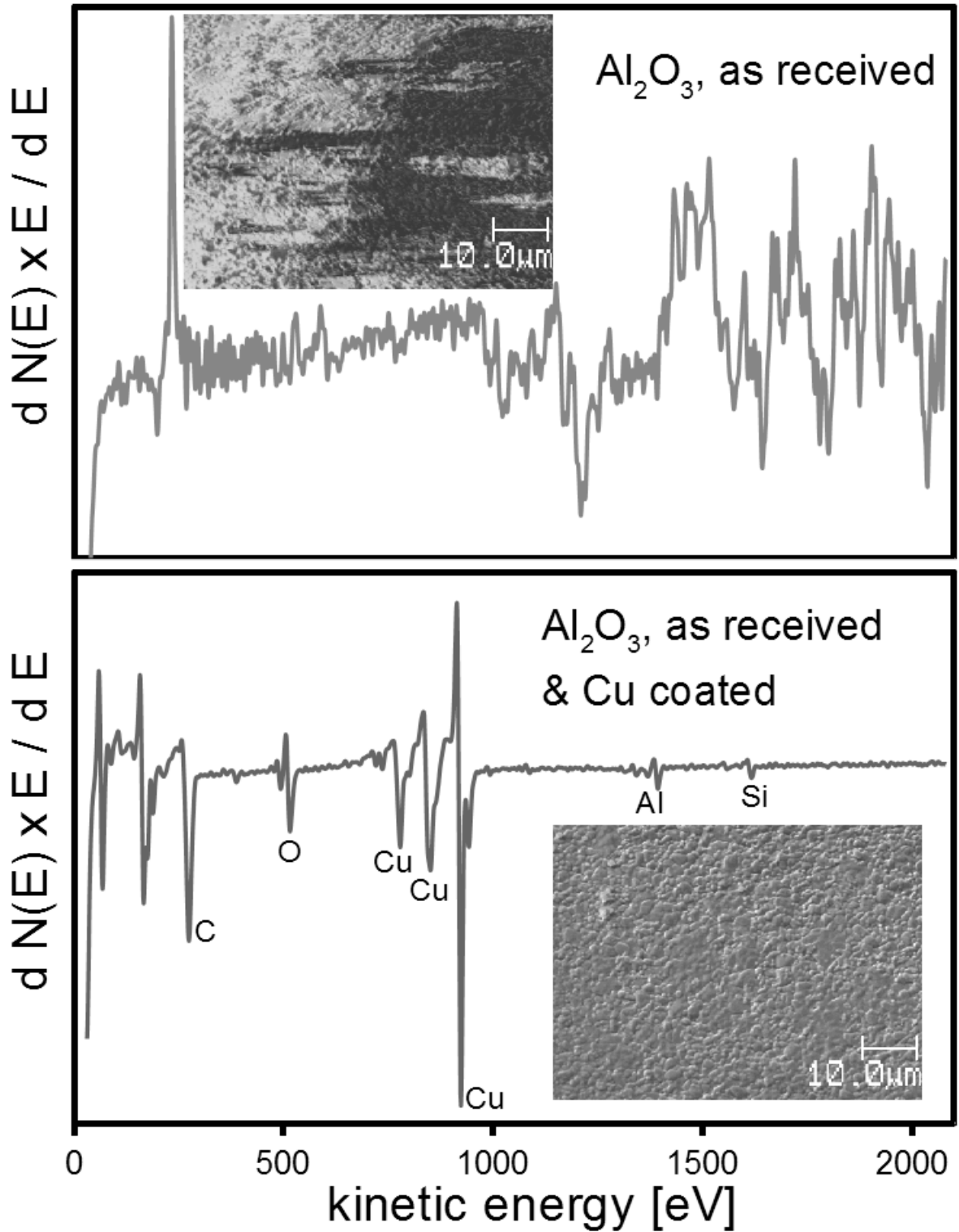

**Fig. 7**:   AES measurements on an insulating, uncoated and coated Al$_2$O$_3$ sample

   An ultra-thin electrical conductive Cu layer is deposit by Reflective Sputtering.

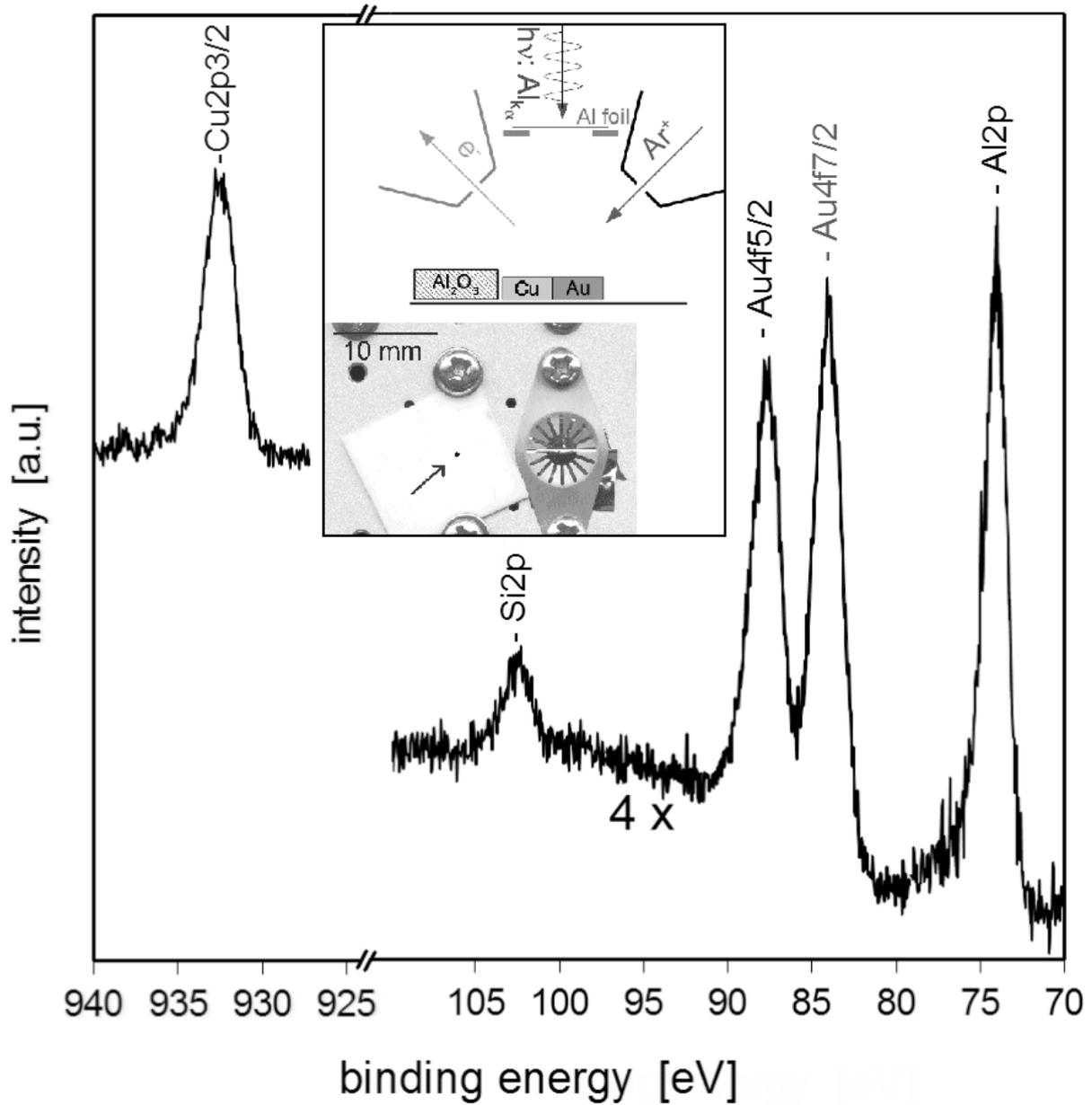

**Fig. 8**:   some peaks of a high energy resolution XPS measurement of an $Al_2O_3$ sample

The sample is coated with Cu and Au in the sub-monolayer range for energy referencing purpose. The experimental set-up is shown in the insert. The position of the measurement area is marked.

|         | binding energy [eV] |          | at%  |
|---------|---------------------|----------|------|
| Al2p    | *74.12*             | 74.06    | 6.7  |
| Au4f7/2 | *84.02*             | **83.96**| 0.48 |
| Au4f5/2 | *87.69*             | 87.64    |      |
| Si2p    | *102.51*            | 102.46   | 0.7  |
| S2p     | *168.73*            | 168.68   | 0.65 |
| Cl2p    | *199.13*            | 199.09   | 0.49 |
| C1s, C-C| *284.78*            | 284.75   | 69.8 |
| O1s     | *531.63*            | 531.64   | 18.9 |
| Cu2p3/2 | *932.56*            | **932.62**| 1.2 |
| Na1s    | *1071.79*           | 1071.87  | 1    |

**Tab. 1**:  peak binding energies after energy-scale correction